\newif\ifUsingACM
\UsingACMfalse
\UsingACMtrue

\ifUsingACM

    \documentclass[sigconf]{acmart}
    \usepackage{balance} 
    \usepackage{algorithm2e}
    \usepackage{algorithmic} 

\else
    \documentclass[conference]{IEEEtran}
    \usepackage{cite}
    \usepackage{graphicx}
    \usepackage{amsmath,amssymb}
    \usepackage{booktabs}
    \usepackage{url}
    \usepackage{hyperref}
    \hypersetup{hidelinks}
    \usepackage{float}
    \usepackage{array}
    \usepackage{balance}
    
\fi

\begin{document}

\ifUsingACM
    \begin{abstract}

Introduced by Juels and Rivest in 2013~\cite{juels2013Honeywords}, Honeywords, which are decoy passwords stored alongside a real password, appear to be a proactive method to help detect password credentials misuse. However, despite over a decade of research, this technique has not been adopted by major authentication platforms. This position paper argues that the core concept of Honeywords has potential but requires more research on issues such as flatness, integration, and reliability, in order to be a practical deployable solution.
This paper examines the current work on Honeyword generation, attacker modeling, and honeychecker architecture, analyzing the subproblems that have been addressed and ongoing issues that prevent this system from being more widely used. The paper then suggests a deployable framework that combines the attacker-resilient, context-aware decoy creation that Honeywords provide with easy integration into existing systems. Honeywords will only move from an academic idea to a practical security tool if technical advances are paired with secure and straightforward architectures, along with adaptive response handling and detailed configuration checks.

\end{abstract}


\ifUsingACM
\fi
    
\title{Advancing Honeywords for Real-World Authentication Security}

\ifUsingACM
\fi

\ifUsingACM

    
    \author{Sudiksha Das}
    \affiliation{%
      \institution{Department of Electrical and\\Computer Engineering \\ Georgia Institute of Technology}
      \city{Atlanta, GA}
      \country{USA}
    }
    \email{sdas410@gatech.edu}
    
    \author{Ashish Kundu}
    \affiliation{%
      \institution{Cisco Research}
      \city{San Jose}
      \state{CA}
      \country{USA}
    }
    \email{ashkundu@cisco.com}

\else 

    \author{
    \IEEEauthorblockN{Sudiksha Das}
    \IEEEauthorblockA{Department of Electrical and\\Computer Engineering\\
    Georgia Institute of Technology\\
    Atlanta, Georgia\\
    sdas410@gatech.edu}
    \and
    \IEEEauthorblockN{Ashish Kundu}
    \IEEEauthorblockA{Cisco Research\\
    San Jose, USA\\
    ashkundu@cisco.com}
    }    

\fi

\maketitle

\fi 

\section{Introduction}
Passwords are the main way our online identity and personal accounts are protected.  Despite being a powerful and important security method, thousands of breaches take place every year. In 2024, organizations reported 4,876 breach incidents, a 22\% jump over 2023, and over 1.7 billion individual breach notifications were sent to affected users~\cite{kiteworks2024breach}. In June 2025, a massive leak of 16 billion credentials, including passwords and session tokens, impacted platforms such as Apple, Google, Facebook, and GitHub~\cite{light2025breach}. Although improvements have been made through techniques such as Multi-Factor Authentication (MFA) and Anomaly Detection, attackers also gain stronger methods to breach systems every day.

Honeywords, which are decoy but realistic-looking passwords stored alongside real ones, offer a unique password defense idea. The system works by saving both the real password and multiple decoys in the database, and a separate, very small system (known as the honeychecker) keeps track of which one of those entries is the real password. When a Honeyword is used during log-in, a separate verification service known as the honeychecker identifies the use of the decoy and can trigger a security alert, helping detect credential misuse at the point of authentication. In theory, this technique allows for active breach detection even in cases where an attacker has access to valid credentials.

The position advanced in this paper is that Honeywords are a theoretically powerful but unused approach to authentication security outside of academia. The likelihood of Honeywords to actually be deployed and function reliably in real authentication systems depends on progress in four key areas: the development of generation techniques that are resilient to realistic attacker models, seamless integration with modern authentication ecosystems such as OAuth2, SAML, OpenID Connect, and MFA frameworks, adaptive response mechanisms capable of maximizing detection rates while minimizing false positives, and ensuring password policies are correct and consistent so Honeywords do not stand out under live system conditions.

\section{The Problem: Authentication Weaknesses and Breach Detection}
Many current password storing frameworks and web authentication platforms rely on secure password hashing functions such as Argon2, bcrypt, or scrypt~\cite{argon2,provos1999future,percival2009scrypt}. These algorithms are designed to require significant time and memory to compute, achieved through memory hardness, which forces attackers to use large amounts of RAM for each password guess, or through high iteration counts, which make every hash calculation slower. Either way, these methods make it more difficult for an attacker to recover passwords from stolen hashes. 

\subsection{Password Hashing Functions and  Crack Efficiency}
The strength and effectiveness of a hashing algorithm is based on how much it slows attackers down per guess, which comes from both RAM requirements and throughput on high-end hardware. The entropy $H$ of a password measures its unpredictability in bits, and each bit of entropy doubles the number of possible passwords. So for a password of length $L$ drawn from an alphabet of size $N$,
\begin{equation}
H \;=\; L \cdot \log_2(N).
\end{equation}
For an 8-character password composed of alphanumeric symbols ($N = 62$), we have $H \approx 47.6$ bits, yielding a total search space of $2^{47.6} \approx 2.9 \times 10^{14}$ combinations. 
Table~\ref{tab:idealcrack} shows the theoretical crack times implied by this full space when combined with benchmarked hash rates.

\begin{table}[H]
\caption{Theoretical exhaustive crack times for a uniformly random 8-character alphanumeric password ($H \approx 47.6$ bits).}
\label{tab:idealcrack}
\centering
\begin{tabular}{@{}lll@{}}
\toprule
Algorithm & Hash Rate (guesses/s) & Exhaustive Time \\
\midrule
Argon2id ($m{=}512$MB, $t{=}3$) & $\sim 60$/s & $\sim$112,700 yrs \\
bcrypt (cost$=12$)              & $\sim 200{,}000$/s & $\sim$34 yrs \\
scrypt ($N{=}2^{15}, r{=}8$)    & $\sim 4,500$/s & $\sim$1,500 yrs \\
\bottomrule
\end{tabular}
\end{table}

\begin{flushleft}
\textit{Sources:} Argon2id rates from \cite{argon2,rfc9106}; bcrypt and scrypt rates from \cite{chick4090gist,hashcatform4090} emperical source.
\end{flushleft}

The exhaustive crack times in Table~\ref{tab:idealcrack} are calculated by combining
the total search space of an 8-character password
($2^{47.6} \approx 2.9\times10^{14}$ candidates) with measured
hash rates. Argon2id throughput ($m{=}512$,MB, $t{=}3$, $p{=}1$) is estimated to be around tens of hashes per second on CPUs~\cite{argon2,rfc9106}, leading to an exhaustive time of around $\sim$112{,}700~years. The parameters $m$ and $t$ control memory allocation and the number of passes over memory, while the optional parallelism
factor $p$ enables limited concurrency at the cost of increased total memory use. GPU benchmarks found bcrypt (cost$=12$), where cost is a logarithmic work factor controlling $2^{\text{cost}}$, at a speed of $\sim$200{,}000~guesses/s~\cite{chick4090gist,hashcatform4090}, leading to an exhaustive search of around $\sim$34~years. Scrypt, configured at $N{=}2^{15}$, $r{=}8$, $p{=}1$, operates around 4{,}500~guesses/s (estimated from forum benchmarks~\cite{chick4090gist,hashcatform4090}), yielding an expected crack
time of $\sim$1{,}500~years. 

These values highlight the theoretical strength of hashing functions under an idealized model where all users select passwords uniformly at random. In this statistical view, the search space is so large that brute-force recovery appears computationally and theoretically impossible, with projected crack times measured in thousands of years.

\subsection{Comparison of Hash Functions in Practice}
In practice, users rarely choose truly random passwords. Instead, they rely on predictable patterns, common words, or personal information. Attackers exploit this behavior by leveraging leaked password datasets, heuristic rules, and optimized cracking tools to shrink the effective search space by several orders of magnitude~\cite{wang2023password}. Rather than attempting all $2.9 \times 10^{14}$ possible combinations, practical attacks often succeed by searching only billions of likely candidates drawn from real-world distributions.

\begin{table}[H]
\caption{Practical crack times for Argon2id, bcrypt, and scrypt under reduced search budgets (billions of guesses).}
\label{tab:hashbench}
\centering
\begin{tabular}{@{}lllll@{}}
\toprule
Algorithm & Parameters & Mem. Cost & Hash Rate & Rel. Time \\
\midrule
Argon2id & $m{=}512$MB, $t{=}3$ & 512MB & $\sim 60$/s & $\sim$2.7 yrs \\
bcrypt   & cost$=12$ & $\sim$4KB & $\sim 200{,}000$/s & $\sim$0.3 d \\
scrypt   & $N{=}2^{15}, r{=}8$ & 16MB & $\sim 4,500$/s & $\sim$13 d \\
\bottomrule
\end{tabular}
\end{table}

\begin{flushleft}
\textit{Note:} Hash rates are drawn from published benchmarks on high-end GPUs (e.g., NVIDIA RTX~4090). 
\end{flushleft}

The practical crack times in Table~\ref{tab:hashbench} are derived by scaling  hash rates to a reduced search budget of $5\times10^9$ guesses (approximately $2^{32}$). For example, at $\sim$60~guesses/s, Argon2id ($m{=}512$MB, $t{=}3$) requires roughly 2.7~years to exhaust this reduced keyspace. In contrast, bcrypt (cost$=12$) achieves around $\sim$200,000~guesses/s, reducing the same search to less than one day, while scrypt ($N{=}2^{15}$, $r{=}8$) completes in about 13~days. These comparisons demonstrate that even under realistic attack models, memory-hard KDFs like Argon2id and scrypt continue to provide impressive crack times. However, the gap between theoretical and practical crack times show how attacker efficiency and user behavior together remove much of the security margin. And as visible through the evidence, password hashing and data breaches still take place at a massive rate.

\subsection{The Detection Gap and Honeyword Motivation}
While these hashing algorithms in the above-section section show their value in making password recovery slower and expensive, they provide no method of detecting password misuse once a password has been breached and hashed. 
Detection of password misuse normally relies on external systems such as behavioral Anomaly Monitoring or Multi-Factor Authentication (MFA) checks, both of which may not catch targeted or low-volume attacks.

\begin{table}[H]
\centering
\footnotesize
\caption{Summary of Honeyword Implementation Details}
\label{tab:honeyImplDetailsTable}
\begin{tabular}{|p{2.5cm}|p{4.8cm}|}
\hline
\textbf{Component} & \textbf{Description} \\
\hline
Generation Methods & Typographical Error, Random String, Corpus-Based (RockYou), PCFG + Markov + PII Hybrid~\cite{erguler2015achieving, bhise2017design, wang2022attacker} \\
\hline
Verification Strategy & Sequential (legacy), Parallel matching and honeychecker notification \\
\hline
Storage Format & Stored alongside real password in hashed form. Linux-style: \texttt{/etc/shadow}, PAM extensions required~\cite{linuxshadow, linuxpam} \\
\hline
Honeychecker & Stateless service maintaining index of the real password per user; isolated via Redis memory store~\cite{genc2018security} \\
\hline
Integration Mode & Middleware plugin in Django, OAuth2 flows; modular APIs required for integration into the identity management systems (IDM) like Keycloak, Auth0~\cite{djangoauth, keycloak} \\
\hline
Detection Outcome & On Honeyword use, the honeychecker triggers adaptive response (MFA, silent logging, account lockout) \\
\hline
\end{tabular}
\end{table}

Honeywords address this limitation by inserting decoy credentials alongside the real password within the password database. If an attacker uses a Honeyword, a separate verification service, or a honeychecker, can flag the attempt, allowing for immediate defensive action to take place. This is an additional beneficial tool not used in the shadow password approaches in Unix/Linux systems~\cite{linuxshadow}, which secure stored credentials but cannot detect the misuse of stolen passwords during login. 

Table \ref{tab:honeyImplDetailsTable} summarizes the main components of Honeyword-based authentication, 
including how decoys are generated, stored, and verified. This overview provides 
a foundation for understanding the detailed attacker models, flatness metrics, 
and deployment challenges discussed in the following sections.

However, the ``flatness'' problem, or ensuring decoys are indistinguishable from the real passwords, is a major issue in practical settings. Simple or naive approaches such as random string generation or simple typographical errors often fail, since attackers can train models on leaked password databases to identify which entries are likely false. That allows the attackers to often successfully filter out implausible decoys using advanced statistical analysis~\cite{chakraborty2022hnyword}. 
However, progress has been made towards improving these decoys through numerous tested methods, which are mentioned in the next section.
\section{Challenges Addressed in Prior Work}

Over the past decade, many academic and industrial research groups have made progress towards some of the technical barriers to Honeyword deployment, especially in the area of decoy generation. 
The main metric that measures the progress of the Honeywords is called $\varepsilon$ flatness. 
This is a measure of how indistinguishable the Honeywords are from the authentic user passwords under rigorous statistical analysis. 
This is a normalized value and the range of this variable can be from $0$ to $1$.
An $\varepsilon$-flatness of $1.0$ indicates perfect indistinguishability, while lower values mean greater vulnerability when under filtering attacks.

\subsection{Attacker Capability Models}
Evaluation results are often reported alongside an attacker capability model with four commonly used levels~\cite{erguler2015achieving,bhise2017design,wang2022attacker}:
\begin{itemize}
\item \textbf{A1} -- Basic dictionary attacker, unaware of of the existence of Honeywords.
\item \textbf{A2} -- Honeyword-aware attacker without access to large external password corpora.
\item \textbf{A3} -- Honeyword-aware attacker with access to public password datasets.
\item \textbf{A4} -- Fully capable attacker with complete awareness of the Honeyword mechanism, access to large datasets, and advanced statistical filtering tools.
\end{itemize}

\subsection{Representative Evaluation Results}
In the year $2015$, Erguler ~\cite{erguler2015achieving} introduced a corpus-based approach. In this technique, to improve the plausibility metric, they used a wide-ranged  dataset from real-world database of password leaks. 
This technique  yielded $\varepsilon$-flatness of $0.85$ and an A4 success rate of $16\%$. 
Later, in the year $2017$, Bhise \& Jagtap ~\cite{bhise2017design} adopted a similar RockYou-based strategy, producing $\varepsilon$-flatness of $0.83$ and an A4 success rate of $19\%$.

\begin{figure}
    \centering
    \includegraphics[width=0.85\linewidth]{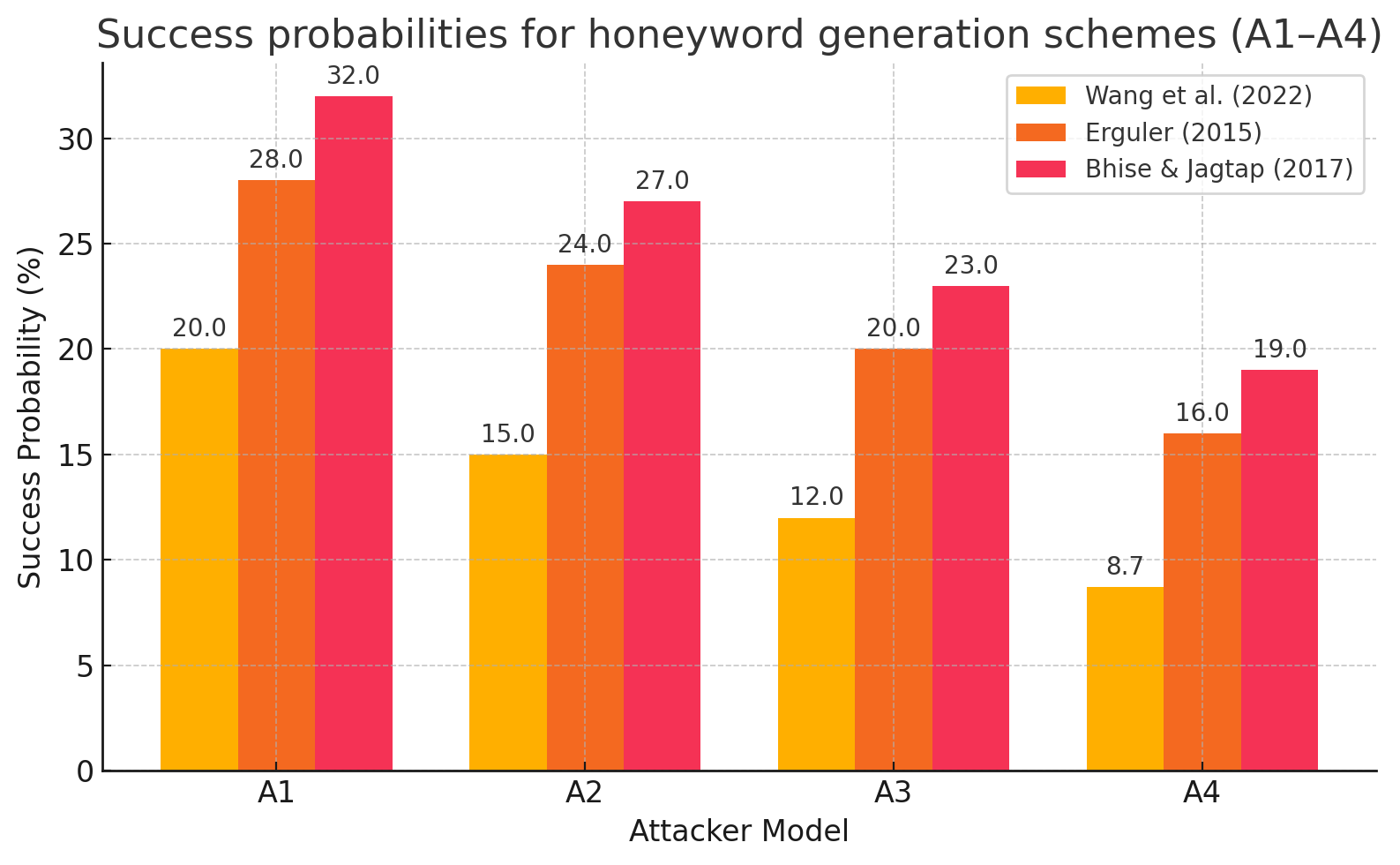}
    \caption{Success probabilities for three Honeyword generation schemes 
    (Wang 2022, Erguler 2015, Bhise \& Jagtap 2017) across attacker models A1--A4. 
    Attacker-aware methods significantly reduce compromise rates, especially in the most capable A4 model. 
    Sources: \cite{erguler2015achieving,wang2022attacker}.}
    \ifUsingACM \Description{Visualization of Attacker Performance} \fi
    \label{fig:attacker_performance}
\end{figure}

The strongest evidence of Honeyword readiness comes from the controlled evaluations such as Wang et al.~\cite{wang2022attacker}. 
Their hybrid PCFG–Markov model, enhanced with targeted PII incorporation, achieved an $\epsilon$-flatness of $0.92$ and reduced attacker success rates from 20\% in the weakest threat model (A1) to just 8.7\% in the most capable (A4). 
By comparison, Erguler’s~\cite{erguler2015achieving} corpus-based method reached $\epsilon = 0.85$ with a 16\% A4 success rate. 
This consistent $5$ to $10$ percentage point advantage across all attacker levels is shown in Figure~\ref{fig:attacker_performance}.

\subsection{Comparative Results}
Table~\ref{tab:flatness_results} summarizes the $\varepsilon$-flatness values and attacker success rates for these representative schemes. Higher $\varepsilon$-flatness correlates with reduced guessability, and this flatness is  essential for the effectiveness of Honeyword approach. The reason is, without it, capable attackers (A3/A4) can filter out the decoys and non-uniform patterns using statistical models, effectively rendering the system’s detection capability pointless. 
If the decoys are easily separable from the real password, perfect integration or operational design won't make a difference, making flatness a key issue in the Honeyword usability~\cite{erguler2015achieving,bhise2017design,wang2022attacker,chakraborty2022hnyword}.

\begin{table}[H]
\caption{Comparative $\varepsilon$-flatness and attacker success rates for representative Honeyword generation schemes under attacker models A1--A4.}
\label{tab:flatness_results}
\centering
\begin{tabular}{@{}llllll@{}}
\toprule
Scheme & $\varepsilon$ & A1 & A2 & A3 & A4 \\
\midrule
Hybrid PCFG + Markov + PII & 0.92 & 20\% & 15\% & 12\% & 8.7\% \\
Corpus-based flatness       & 0.85 & 28\% & 24\% & 20\% & 16\% \\
RockYou-based               & 0.83 & 32\% & 27\% & 23\% & 19\% \\
\bottomrule
\end{tabular}
\end{table}

\textit{Note:} $\varepsilon$-flatness and A4 success rates are reported directly from the cited works~\cite{erguler2015achieving,bhise2017design,wang2022attacker}. 
Values for A1–A3 attacker models are representative estimates based on the descriptions presented in these papers, and illustrate the relative degradation across the attackers' capabilities.

\subsection{Visualization of Attacker Performance}

These studies together demonstrate the technical groundwork for Honeyword-based authentication and measurable gains in flatness, attacker resistance, and honeychecker isolation. However, despite these advances, substantial barriers remain before Honeywords can transition from primarily academic-level prototypes to deployable production-grade authentication defenses. The next section outlines these unresolved challenges, divided into research and industrial engineering domains.

\section{Remaining Research Challenges}\label{sec:Remaining Research Challenges}

In spite of research leading to improved Honeyword generation techniques and stronger honeychecker architectures, the Honeyword mechanism have not led to wide-spread real world implementation. The remaining research challenges should focus less on how Honeywords work in theory, and more about ensuring that they remain robust, realistic, and verifiable under real-world attack models.

\subsection{Flatness and Decoy Realism}
 Even with the most advanced $\varepsilon$-flatness techniques, skilled attackers with access to large password leaks, user-specific PII, or registration metadata may still be able to filter out unlikely passwords using statistical and contextual cues~\cite{erguler2015achieving,bhise2017design,wang2022attacker}. Hybrid PCFG/Markov generation strategies reduce this risk in controlled tests, but there is little evidence they offer the same protection under real-world conditions. Honeywords can still add another layer of protection if implemented carefully, though more research is needed on how well they hold up in current real-world conditions.
 So, future research must focus on more on adaptive, user-aware generation methods that maintain flatness while scaling across diverse linguistic and contextual password patterns.

\subsection{Personalized Generation and False Positives}
Personalization, creating decoys tailored to a user’s habits or language, can help improve the believability of decoy passwords~\cite{chakraborty2022hnyword}. However, naive personalization can backfire: if decoys are too close to the real password, legitimate mistypes or auto-corrects may trigger Honeyword alerts. Usability issues such as typographical errors, multilingual keyboards, and auto-correction further increase the likelihood of false positives, eroding user trust. Research is needed on adaptive personalization that preserves diversity without sacrificing reliability. This raises the probability of false alerts if a legitimate user accidentally enters a decoy, such as in settings with mobile auto-correct. 

\subsection{Honeychecker Robustness}
Beyond generation quality, research has also examined honeychecker robustness. In  the year $2018$, Gen\c{c} et al.\ ~\cite{genc2018security} found that traditional honeychecker systems may be vulnerable to both code corruption and denial-of-service attacks. To solve these challenges, research has proposed architectural compartmentalization, such as a Redis-style isolation, to limit the potential impact of denial-of-service attacks on the verification component. However, there is still not much proof of improved resilience in this new system of honeychecker.

\subsection{Lack of Comparative Studies}
Despite numerous proposals, there are very few studies that test and compare all these proposed Honeyword schemes under identical, rigorous and realistic conditions~\cite{chakraborty2022hnyword}. 
In general, controlled experiments are typically small-scale and rely on synthetic experimental setups. 
Large-scale testing, especially across varied user demographics, password policies, and adversary models would greatly help in accessing the practicality and effectiveness of the Honeyword system. 
Without a solid framework to balance effectiveness with usability, decoy reliability remains an open research problem~\cite{bhise2017design,wang2022attacker}.

\section{Remaining Engineering and Deployment Challenges}\label{sec:Remaining Engineering and Deployment Challenges}
Even though the concept of Honeywords is well-proven and theoretically sound, their deployability depends on how easily they integrate with real authentication ecosystems. The following challenges involve interoperability, organizational adoption, and operational reliability.

\subsection{Integration with Authentication Frameworks}
One of the biggest challenges is that there’s no standard way to plug Honeywords into the frameworks everyone already uses for authentication. Chakraborty et al.\ (2022)~\cite{chakraborty2022hnyword} offered a comprehensive view of the different Honeyword schemes and highlights another persistent issue: despite many proposals, few empirical studies compare competing designs under realistic conditions. Honeywords aren’t built into widely used identity systems like OAuth2, OpenID Connect, SAML, or enterprise MFA platforms. In real-world deployments, these frameworks serve as the backbone of identity verification, so any change has to preserve protocol compliance, keep performance fast, and pass security audits. Without standard middleware or plugin APIs, there’s no straightforward way to support Honeywords across these systems. Teams would need to rework core login flows, wire honeychecker checks into token steps, and make sure any added delay goes unnoticed.

\begin{table}
\caption{Password hashing defaults and supported options in popular frameworks.}
\label{tab:framework_hashers}
\centering
\resizebox{\linewidth}{!}{%
\begin{tabular}{@{}lll@{}}
\toprule
Framework / Platform & Default Hashing Scheme & Other Supported Options \\
\midrule
Django (Python) & PBKDF2 (SHA-256) & Argon2id, bcrypt, PBKDF2-SHA1~\cite{djangoauth, djangoauth2} \\
Ruby on Rails & bcrypt & Argon2 (gem), PBKDF2 (plugin)~\cite{railsguide} \\
Spring Security (Java) & PBKDF2 (SHA-256) & bcrypt, scrypt, Argon2~\cite{springsecurity} \\
ASP.NET Core (.NET) & PBKDF2 (HMAC-SHA1) & bcrypt, scrypt, Argon2~\cite{aspnetpw} \\
Node.js & bcrypt (npm) & scrypt, Argon2 (npm)~\cite{nodecrypto,bcryptnpm} \\
PHP & bcrypt (\texttt{PASSWORD\_BCRYPT}) & Argon2i, Argon2id~\cite{phppass} \\
Go (Golang) & bcrypt & scrypt, Argon2id~\cite{gocrypto} \\
Linux PAM & SHA-512 (\texttt{crypt}) & bcrypt, scrypt, yescrypt, Argon2~\cite{linuxpam} \\
\bottomrule
\end{tabular}%
}
\end{table}

One way to understand the integration challenge is to look at the password hashing functions currently supported across common frameworks. 
Although Argon2id has is the academically most common and recommended choice, most deployed systems continue to default to PBKDF2 or bcrypt. 
Table~\ref{tab:framework_hashers} summarizes these defaults, illustrating the ecosystem reality that Honeywords must adapt to in order to be deployable.

\subsection{Adoption Barriers Compared to Existing Defenses}
Compared to the traditional defenses like Anomaly scoring, Device fingerprinting, and Geofencing~\cite{chakraborty2022hnyword} etc, Honeyword-approach is significantly less known in the community. These above-mentioned, more-popular methods are already built into day-to-day operations, come with clear playbooks for handling alerts, and are easy to manage with existing tools. Large identity providers and enterprise teams are familiar with these defenses, while Honeywords lack a standardized integration path. The mix of risk and cost makes adoption unrealistic, even if the underlying algorithms are technically proven~\cite{erguler2015achieving,wang2022attacker,djangoauth, djangoauth2, keycloak}.

\subsection{Policy and Configuration Reliability}
A final challenge is with the password policies and configuration reliability. Honeywords can provide efficient detection-value only if the real password and all the decoys comply with the password policies, ensuring that they are all accepted and appear plausible. 
In one direction, misconfigurations that allow too-simple decoys are easy to breach into.
On the other hand, too strict configurations that rejects realistic decoys, can also silently weaken the effectiveness of this approach. 
On Linux systems, for example, rules are enforced through the Pluggable Authentication Modules (PAM) component \texttt{libpwquality}, configured via \text{pwquality.conf}. 
Recent work by Vaidya et al. ~\cite{vaidya2025reliability} demonstrated that even advanced Large Language Models (LLMs) often generate \text{pwquality.conf} files that are incomplete, inconsistent, or contain hallucinated parameters, which could silently weaken policy enforcement or interfere with honeychecker operation. 
This underscores the need for automated validation and robust integration at the policy layer, since fragile configurations can undermine the very detection guarantees that Honeywords are meant to provide.

\subsection{Summary of Deployment Barriers}

In summary, while research has made progress on things like flatness and honeychecker security, the main challenges now lie in the integration, institutional adoption, resilience against targeted attacks, personalization trade-offs, and configuration reliability. Overcoming these will require coordinated progress in system architecture, attacker-aware decoy modeling, and automated policy verification.

\section{Proposed Approaches to Remaining Challenges}
Given the deployment barriers outlined in Section \ref{sec:Remaining Engineering and Deployment Challenges}, a proposed future solution focuses on developing an efficient implementation-oriented framework. That mechanism needs to address integration, personalization, response-time, and configuration reliability in a way that can be refined through more aggressive and holistic testing methodologies.

\subsection{Middleware for Integration}
The integration gap can be closed through creation of a modular middleware that can plug directly into popular authentication frameworks like Django, Spring Security, or Keycloak~\cite{djangoauth2,keycloak}. The role of this middleware is  to intercept all the login attempts, to validate the submitted credentials against both the password database and the honeychecker, and then integrate those results into existing Risk-based Authentication or Anomaly Detection systems. To keep everything secure, honeychecker operations should be isolated, using a Redis-backed compartmentalization model~\cite{genc2018security}, so that the main authentication service never has direct access to the real password index. This reduces the attack surface and makes the system more fault-tolerant. Additionally, with this kind of middleware, Honeywords can be implemented into existing systems without major rewrites.

\subsection{Authentication Flow and Honeychecker Isolation}
\begin{figure*}
    \centering
    \includegraphics[width=0.9\linewidth]{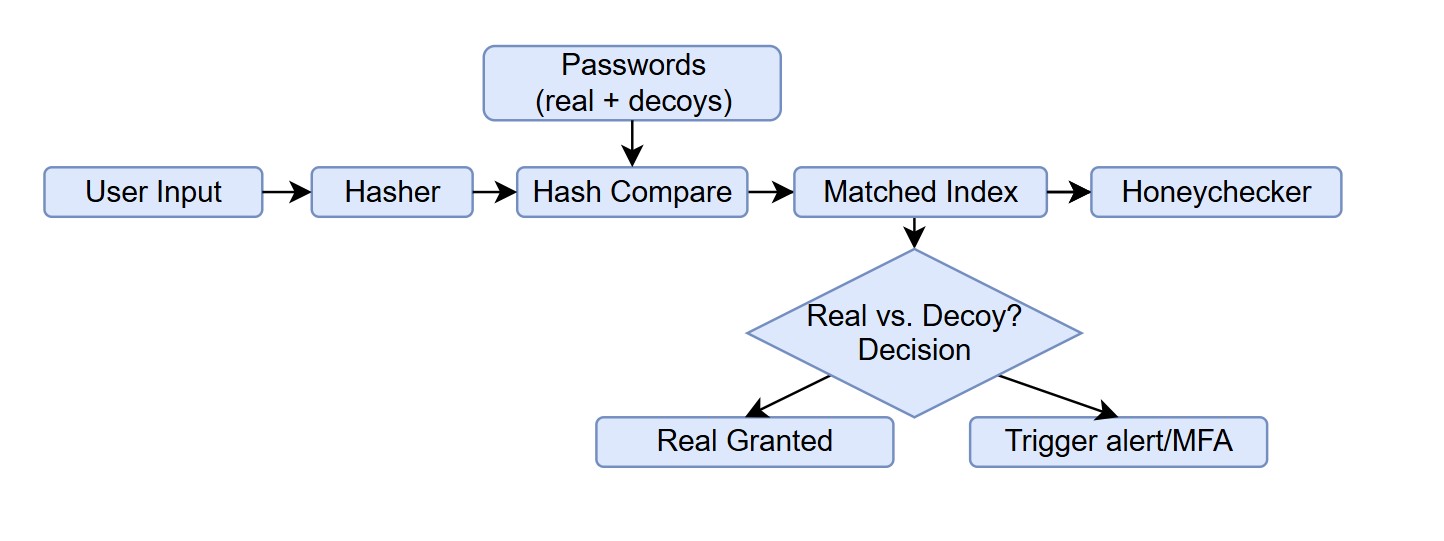}
    \caption{Proposed Honeyword-augmented authentication architecture that integrates seamlessly into Django and OAuth2 flows via framework-specific hooks (ex. Django custom authentication backends, OAuth2 pre-token issuance plugins).}
    \ifUsingACM \Description{Proposed Honeyword-augmented authentication architecture} \fi
    \label{fig:Honeyword_augmented_authentication_architecture}
\end{figure*}

This system design begins with a user attempting to to perform a log in operation. First, this request  passes through the standard authentication mechanism, where the submitted password is hashed using the targeted site’s configured algorithm (ex. Argon2id, bcrypt, or scrypt). After that, this is  compared against all the stored Honeyword hashes for that specific account. 
Figure \ref{fig:Honeyword_augmented_authentication_architecture} illustrates that the Honeyword verification occurs in parallel with the password comparison, ensuring minimal delay and immediate detection. This is more efficient compared to the earlier sequential models, where verification occurred only after completion of the user authentication process. Minimal delay is important, so that the potential unauthorized user does not get a hint of being caught. If a match is found, the authentication middleware notifies the isolated honeychecker with the associated Honeyword index. The honeychecker, which maintains only an in-memory mapping of usernames to the index of the true password, returns either “Real” or “Honeyword” flag~\cite{juels2013Honeywords}.

A “Real” result means that the entered password is correct, which allows the remaining systems to continue. On the other hand, a “Honeyword” result initiates the adaptive response module, which can trigger escalated security measures such as silent logging, Multi-Factor Authentication (MFA) prompts, issuing restricted-scope tokens, or even locking the account. The isolation of the Honeyword checker ensures that the mapping of true password indices is never stored on disk, minimizing the risk of the of a successful hack in the event of a system breach. Additionally, the idea behind the design of these integration points is to require minimal changes to existing protocol-level logic, allowing Honeyword detection to work as a drop-in enhancement without disrupting existing identity workflows~\cite{genc2018security}.

\subsection{Enhancing Decoy Generation}
The next priority in this integration layer is to enhance decoy generation to make sure that the set of Honeywords are both believable and policy-compliant. This will involve adapting hybrid PCFG/Markov models~\cite{wang2022attacker} discussed earlier to match the password policies and composition rules of the target environment. Additional work should be completed so that this generation can incorporate user-specific behavioral cues such as keyboard layout, typing frequency, and password length preferences. These enhancements need to ensure that there is minimal amount of exploitable statistical biases or there is no  increase in false-positive rates. Earlier corpus-based approaches~\cite{erguler2015achieving,bhise2017design} and survey findings~\cite{chakraborty2022hnyword} provide baselines that highlight both the progress and limits of decoy generation. More research experiments need to be done on the pros and cons of current decoy generation methods, to find the best one. In this competitive evaluation process, it would be possible to  include  applications of LLMs, agents, or generation of Honeywords from tree structures.

\subsection{Adaptive Response Mechanisms}
Adaptive response mechanisms should include reactions more than "allow or block" for that attempted login. In this mechanism, the backend system should silently trigger some additional reactions, even leading to an escalation over time, if needed. By factoring in IP reputation, geo-location anomalies, device fingerprint mismatches, and historical login context, the system can trigger graduated responses. These responses typically start from silent logging and monitoring. Based on the analysis of the logged-data, this process can later initiate  stepped-up authentication challenges through some additional confirmation, such as one-time password (OTP), re-verification of previously-set authentication questions, push notification in the mobile device's App etc. If the suspicion continues to rise, this process can eventually trigger a temporary account lockdown. In this way, the Adaprove Response Mechanism  enables early breach-detection while remaining operational and functional~\cite{light2025breach,kiteworks2024breach}.

\subsection{Path Towards Deployment}
This proposal is intended as a starting point for implementation. By first deploying the middleware and honeychecker in a controlled environment, we can measure detection rates, false positives, and operational impact under real-world conditions. These results will guide iterative refinements to decoy generation, personalization strategies, and adaptive response policies, ultimately bridging the gap between research prototypes and production-grade Honeyword deployments~\cite{shadmand2025passwords,argon2,rfc9106,provos1999future,percival2009scrypt}.

\section{Related Work}

The concept of Honeywords occupies a distinct and well-understood role in the authentication domain of Cybersecurity. They detect the misuse of valid credentials during the time of login. Unlike most defense techniques, they don’t just make compromise harder, they also issue alert-signal when compromise has happened.

\subsection{Patents on Honeywords}

Honeywords was first introduced by the  academic world \cite{juels2013Honeywords}. Over the years, the idea has gained some traction from professional users, as shown through a number of patents. 
As of year $2025$, at least $39$ patents address Honeyword-based authentication or related deception mechanisms, with noticeable growth in patent-filings starting around the year $2019$. Several notable patent examples include the following:

\begin{itemize}
    \item IBM’s $2017$ patent \cite{us9584507} introduces a distributed authentication system that separates password verification process across entities, improving compromise detection through isolation.
    
    \item NortonLifeLock’s  designed process \cite{us11438378} in $2022$ completely hides the presence of Honeywords during the login validation process, strengthening the defense against the side-channel analysis.

    \item EMC’s $2017$ patent \cite{us9843574} defines methods for generating both synthetic and non-synthetic “chaff passwords,” a concept closely related to Honeywords, and proposes deployment in hardened password systems.

    \item Symantec's encryption scheme \cite{us9325499} explores the use of “honeymessages” during the honey encryption. These messages efficiently leverages low-entropy keys in the adversarial environments.
    
\end{itemize}

The system described in this paper aligns with several core themes in the existing Honeyword patents, while expanding them through middleware compatibility, adaptive response logic, and modular deployment.

\subsection{Comparison to Established Defenses}
Well-established techniques of Password hashing, Multi-Factor Authentication (MFA), Anomaly Detection etc. help in many different ways. 
A recent study found that Multi-Factor Authentication cuts the risk of compromise by over $99$\%, even for users with exposed credentials~\cite{meyer2023mfa}. Still, Multi-Factor Authentication can be bypassed via phishing or session hijacks techniques. 
As for Anomaly Detection, industry data shows that most organizations ($46$\%) only detect attacks after the targeted accounts are already compromised, and a further $27$\% detect them even later~\cite{abnormal2024mfa}. 
Other decoy-based tactics like honeypots work at the network level, but are not part of the user authentication step, during the login time. 
In contrast, Honeywords approach embed detection directly into the login authentication step, offering an immediate alert if a compromised or stolen password is used.

\subsection{Conceptual Comparison}
\begin{table}
\caption{Conceptual comparison of detection capabilities.}
\label{tab:defense_conceptual}
\resizebox{\linewidth}{!}{%
\begin{tabular}{@{}llll@{}}
\toprule
Mechanism & Detection Target & Relative Detection & Relative False Positives \\ 
\midrule
Honeywords & Credential misuse at login & High (needs $\varepsilon$-flatness $\geq 0.9$) & Moderate \\
MFA & Suspicious logins & High & Low \\
Anomaly Detection & Account takeover & Moderate & Moderate \\
\bottomrule
\end{tabular}%
}
\end{table}

Conceptual comparison of different detection capabilities are presented in Table~\ref{tab:defense_conceptual}. 
For that data-set, Honeyword stats was collected from \cite{erguler2015achieving,wang2022attacker},  
Multi-Factor Authentication data was collected from Meyer et al.~\cite{meyer2023mfa} and the
Anomaly Detection information came from the industrial data~\cite{abnormal2024mfa}.

False positive rates of the Honeyword approach may be somewhat higher than those of mature Multi-Factor Authentication or Anomaly Detection systems. That highlights the need for careful tuning and usability safeguards in the Honeyword mechanism. At the same time, their unique advantage is that they provide direct, verifiable evidence of credential misuse at the point of login. Unlike Multi-Factor Authentication or Anomaly Detection, which infer compromise indirectly, Honeywords confirm it when it happens. This makes them a valuable complement to existing defenses and underscores their potential role in strengthening today’s production identity systems.
\section{Discussion}

Honeywords are at a stage where the question is no longer whether they work, but how to improve their chances of decoy success and how to deploy them effectively in real authentication systems. 
As discussed in earlier sections, their advantage is unique: they detect the misuse of valid credentials right at the moment when it is happening, something neither Multi-Factor Authentication nor Anomaly Detection can guarantee.

\subsection{From Theory to Deployment}

\begin{figure}
    \centering
    \includegraphics[width=0.85\linewidth]{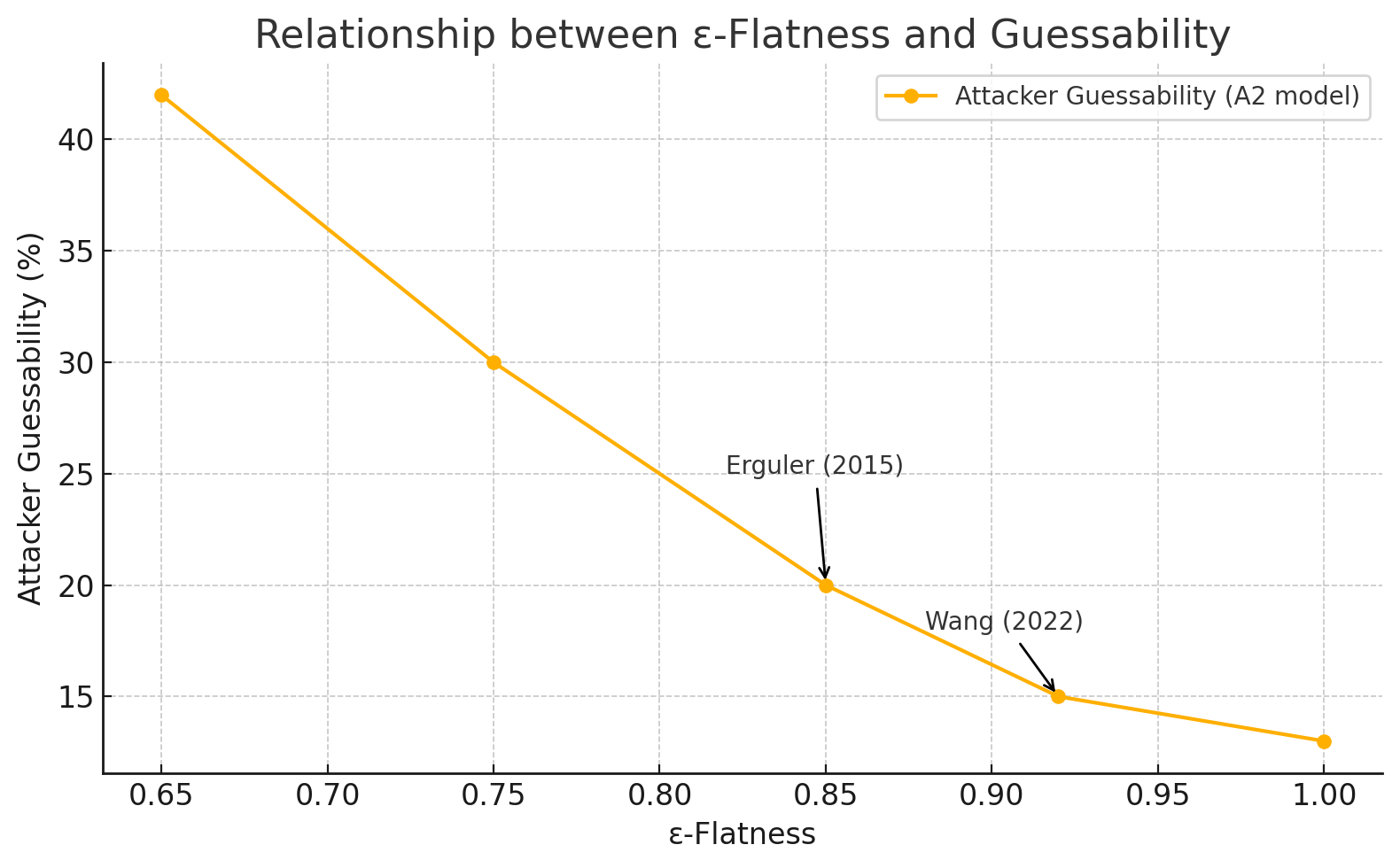}
    \caption{Relationship between $\epsilon$-flatness and attacker guessability for an A2-level attacker. Higher flatness reduces guessability, but improvements beyond $\epsilon=0.9$ yield diminishing returns, suggesting that deployment focus should shift to integration and adaptive detection at that point. Sources: \cite{erguler2015achieving,wang2022attacker}.}
    \ifUsingACM
    \Description{Relationship}
    \fi
    \label{fig:flatness_curve}
\end{figure}

While high $\epsilon$-flatness is necessary to resist statistical filtering, gains diminish beyond roughly $\epsilon = 0.9$. In Wang et al.’s A2-level results, raising flatness from $0.85$ to $0.92$ lowered the guessability metric by about $5$ percentage points. Further increases toward 1.0 would yield much smaller improvements. Figure~\ref{fig:flatness_curve} illustrates this curve showing that resistance rises sharply until about $\epsilon = 0.85$, and then it somewhat levels off.
Published studies report discrete points, ex. Erguler (0.85, 16\% A4), Bhise (0.83, 19\%), Wang (0.92, 8.7\%). The trend suggests diminishing gains beyond $\varepsilon \approx 0.9$, but more empirical work is needed to confirm this.

\subsection{Position Statement}

Two points central to this paper’s position: (1) advanced generation techniques can meaningfully reduce attacker effectiveness even under strong threat models, and (2) the primary gap is deployment, as no published evidence yet demonstrates equivalent performance in live, large-scale systems.

The combined takeaway is that the core technical challenge of generating realistic, attacker-resilient Honeywords, effective even against strong A4-level adversaries has made substantial progress, though more work should be done for personalization of Honeywords
. The primary remaining research challenge is ensuring these high-flatness schemes can be integrated into live authentication ecosystems while preserving usability, interoperability, and detection reliability.

\section{Questions for the Community}
\label{sec:community}

The transition from useful academic prototypes to production-ready systems raises a new set of interdisciplinary, open-ended questions that must be explored across industry, academia, and standards bodies.

\subsection{Open Discussion Points}

\begin{enumerate}
    \item \textbf{What would it take for real-world adoption?} Why have Honeywords not been widely deployed, despite strong theoretical backing? Are the barriers primarily technical (ex. middleware integration), organizational (ex. risk management), or perceptual (ex. lack of awareness)? Could there be some platform-specific pilot studies that would help validate the feasibility of this concept?

    \item \textbf{Can existing identity platforms evolve to support Honeywords?} What specific changes would the existing well-known platforms like Azure AD, Auth0, Okta, or Keycloak need to do, to successfully adopt decoy-based authentication? Would middleware integration suffice~\cite{genc2018security, djangoauth2}, or does that require an extensive revision of the standards like OAuth2 and OpenID?

    \item \textbf{How do Honeywords fit within a passwordless future?} As FIDO2, passkeys, and biometrics slowly displace traditional passwords, is there still a role for Honeywords as a post-compromise signal? Could hybrid systems enable fallback detection using decoys?

    \item \textbf{Should client tools support Honeywords?} Should password managers and browsers support Honeyword generation or detection mechanism as a built-in feature? Could autofill or form-typing behaviors introduce privacy risks or usability trade-offs?

    \item \textbf{What are best practices for response and alerting?} What actions should systems take when a Honeyword is used? Should they opt for silent monitoring, Multi-Factor Authentication prompts, session restrictions, or full account lockout? How can these responses be tuned to balance security aspect with  usability~\cite{light2025breach, abnormal2024mfa}?

    \item \textbf{What explains adoption inertia?} Why have the major commercial cloud identity providers not yet embraced Honeywords~\cite{chakraborty2022hnyword}? Is it due to unclear Return on investment, absence of standards, fear of false positives, or a lack of demand from enterprise security teams?
    
\end{enumerate}

\section{Conclusion}

Honeywords bring something different to the table in authentication security: they can directly flag when stolen credentials are being used. That kind of real-time misuse detection is still missing from most defenses we rely on today. Over the last decade, research has made real progress in generating realistic decoys and designing stronger honeycheckers, but almost none of these carried over into live systems. The gap comes down to integration headaches, usability concerns, and the lack of solid policy checks to make sure decoys actually hold up under real conditions.

Moving forward, what is needed is less theory and more work on deployment. That means figuring out standard ways to plug Honeywords into common frameworks like OAuth2, OpenID Connect, and MFA; building smarter response mechanisms that can adapt based on context instead of just locking users out; and adding automated tools to check that both real and decoy passwords follow the same rules and remain indistinguishable in practice. We see this as the real open research problem: testing Honeywords in production-like environments to measure performance, false positives, and resilience under real attack conditions.

The framework outlined in this paper, covering realistic decoy generation, isolated honeychecker design, adaptive risk scoring, and configuration validation gives a starting point for that transition. With focused research and practical engineering, Honeywords could move from being an academic idea that “works on paper” to a deployable control that strengthens modern authentication systems against credential misuse.

\balance

\end{document}